\documentclass[12pt,reqno,a4paper]{amsart}
\usepackage{amsmath,amssymb,dsfont,graphicx,cite,latexsym,
epsf,cancel,epic,eepic}
\usepackage[colorlinks=false]{hyperref}
\usepackage{mathpazo}
\newtheorem{theorem}{Theorem}

\newtheorem{lemma}{Lemma}

\def\beq{\begin{equation}}
\def\eeq{\end{equation}}
\def\bea{\begin{eqnarray}}
\def\eea{\end{eqnarray}}
\def\benpf{\noindent {\textbf{{\emph{Proof.}}\;}}}
\def\endpf{\hfill$\blacksquare$\medskip}
\makeatletter
\expandafter\let\expandafter
\reset@font\csname reset@font\endcsname
\def\subeqnarray{\arraycolsep1pt
   \def\@eqnnum\stepcounter##1{\stepcounter{subequation}
       {\reset@font\rm(\theequation\alph{subequation})}}
\jot5mm     \eqnarray}

\makeatother

\def\epsilon{\varepsilon}
\def\t{\widetilde}

\def\nn{\nonumber}
\def\endpf{\hfill$\square$\medskip}

\newbox\meibox
\def\placeunder#1#2#3#4{\setbox\meibox%
\vbox{\hbox{\hskip#4$\hphantom{#2}$}\hbox{$\hphantom{#1}$}}%
\vtop{\baselineskip=0pt\lineskiplimit=\baselineskip%
\lineskip=#3\hbox to \wd\meibox{\hfil\hskip#4$#2$\hfil}%
\hbox to \wd\meibox{\hfil$#1$\hfil}}}
\def\intprod{\mathbin{\hbox to 6pt{%
                 \vrule height0.4pt width5pt depth0pt
                 \kern-.4pt
                 \vrule height6pt width0.4pt depth0pt\hss}}}

\begin{document}
\title[On the classification of multidimensionally consistent 3D maps]
{On the classification of multidimensionally consistent 3D maps}

\author{Matteo Petrera \and Yuri B. Suris }

\thanks{E-mail: {\tt  petrera@math.tu-berlin.de, suris@math.tu-berlin.de}}

\maketitle

\begin{center}
{\footnotesize{
Institut f\"ur Mathematik, MA 7-2\\
Technische Universit\"at Berlin, Str. des 17. Juni 136,
10623 Berlin, Germany
}}
\end{center}

\begin{abstract} We classify multidimensionally consistent maps given by (formal or convergent) series of the following kind:
$$
T_k x_{ij}=x_{ij} + \sum_{m=2}^\infty A_{ij ; \, k}^{(m)}(x_{ij},x_{ik},x_{jk}),
$$
where $A_{ij;\, k}^{(m)}$ are homogeneous polynomials of degree $m$ of their respective arguments. The result of our classification is that the only non-trivial multidimensionally consistent map in this class is given by the well known symmetric discrete Darboux system
$$
T_k x_{ij}=\frac{x_{ij}+x_{ik}x_{jk}}{\sqrt{1-x_{ik}^2}\sqrt{1-x_{jk}^2}}.
$$

\end{abstract}


\section{Introduction}

The goal of this paper is to contribute to the problem of classification of discrete three-dimensional (3D) integrable systems.

The notion of integrability we adhere to is the \emph{multidimensional consistency}. It originates in the development of integrable structures of Discrete Differential Geometry, see \cite{DDG} for details and for historical remarks. Multidimensional consistency of a given discrete $d$-dimensional system means that it can be imposed in a consistent way on all $d$-dimensional sublattices of a $(d+1)$-dimensionl lattice. In retrospect, this property can be understood as a discrete analog of the existence of commuting hierarchies of integrable systems. It has been shown that this property guarantees the existence of such integrability attributes as the discrete zero curvature representation and B\"acklund transformations enjoying Bianchi-type permutability. Moreover, these attributes are encoded in the system itself and can be found in a straightforward and algorithmic way. Thus, multidimensional consistency can be accepted as a rather general and powerful definition of integrability of discrete systems. 

In Ref. \cite{ABS}, a classification of 2D systems on quad-graphs with scalar (complex) fields at vertices, integrable in the sense of 3D consistency, was performed. As a result of this classification, a rather short but exhaustive list of integrable systems of this kind was produced, known nowadays as the ABS list. Moreover, one class of multidimensionally consistent 3D systems has been classified, namely those of the octahedron  type \cite{ABS2}. It was demonstrated that the relevant combinatorial structure in this case is a multidimensional root lattice of type A. The most striking feature is that the number of integrable systems drops dramatically with increasing dimension: only half a dozen of discrete 3D systems with the property of 4D consistency are known, almost all of them are of a geometric origin.

In the present paper, we classify 3D maps within certain general Ansatz, which are 4D consistent. The main result of our classification is that, within the considered class, the only non-trivial 4D consistent map is the symmetric discrete Darboux system (equation \eqref{eq: Darboux shift} below). Note that this system appeared in the study of multi-dimensional circular nets \cite{KSch}, see also \cite{BMS}, and enjoys several distinct geometric interpretations, for which one can consult \cite{DDG} and \cite{PS}.

\section{The 4D consistency property}

Consider a 3D map $\Phi(x,y,z)=(\t x_,\t y,\t z)$. A possible definition of integrability of such maps is their 4D consistency \cite{DDG}, which we describe now.

Combinatorially assign the quantities $x=x_{23}$, $y=x_{13}$, $z=x_{12}$ to the three faces of a 3D cube parallel to the coordinate planes $23$, $13$, $12$, respectively. Let $T_k$ stand for the unit shift in the $k$-th coordinate direction. We assign the
quantities $\t x_{ij}=T_k x_{ij}$  to the three opposite faces. Therefore we view the map $\Phi$ as
\beq \label{eq: map}
\Phi_{123}\left(x_{12},x_{13},x_{23}\right)=(T_3 x_{12},T_2 x_{13},T_1 x_{23}),
\eeq
where the indices $123$ of $\Phi$ represent the 3D space where the map acts. More generally, we assume that in any 3D sublattice of $\mathbb Z^N$ spanned by the directions $i,j,k$ $(i < j <k)$ there acts a map $\Phi_{ijk}$.
\begin{figure}[htbp]
\begin{center}
\includegraphics[height=6cm]{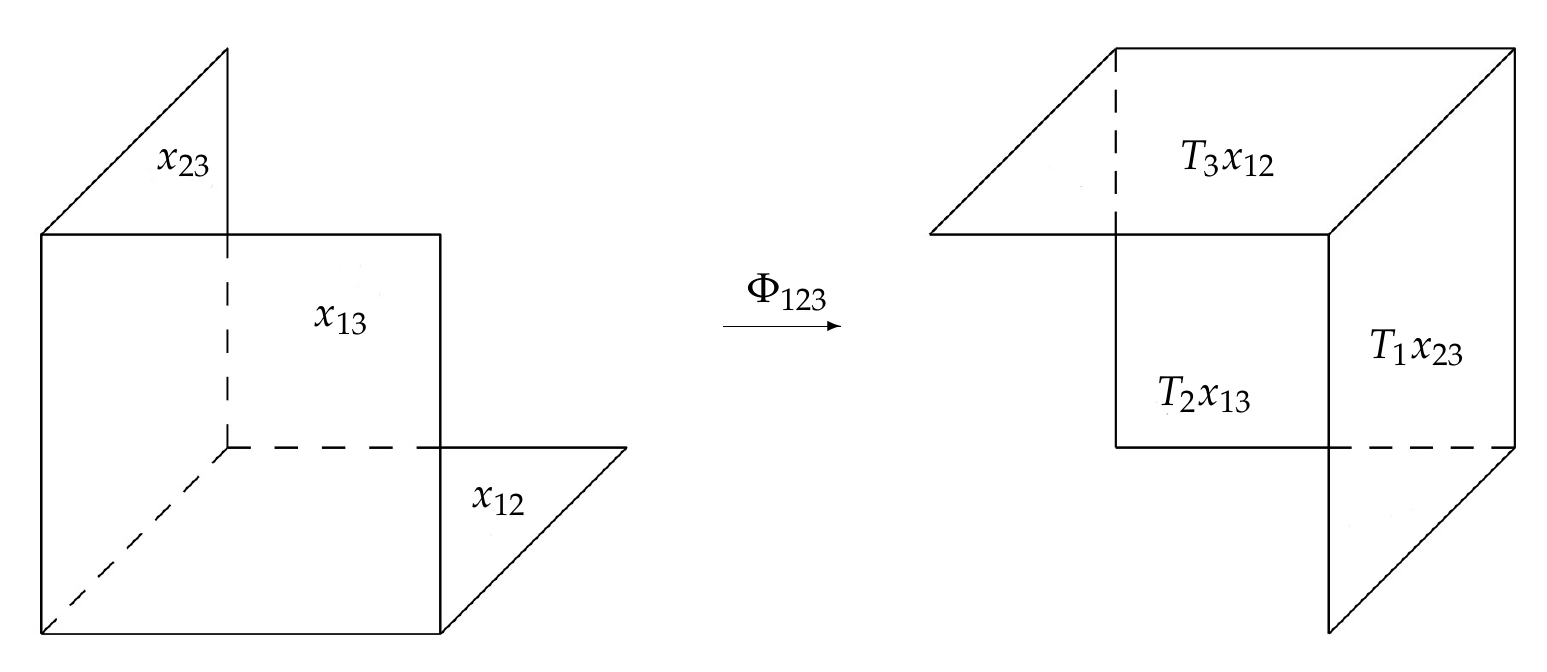}
\caption{A map on an elementary 3D cube with fields assigned to 2-faces.
}
\label{Fig:cube eq faces}
\end{center}
\end{figure}

\noindent Now consider the initial value problem with data $x_{ij}$, $i,j=1,2,3,4$, $i<j$, prescribed at six squares adjacent to one common vertex of the 4D cube. Then the application of a 3D map like (\ref{eq: map}) to the four 3D cubes adjacent to this vertex allows us to determine all $T_k x_{ij}$:
\begin{eqnarray*}
&& \Phi_{123}\left(x_{12},x_{13},x_{23}\right)=(T_3 x_{12},T_2 x_{13},T_1 x_{23}), \\
&& \Phi_{124}\left(x_{12},x_{14},x_{24}\right)=(T_4 x_{12},T_2 x_{14},T_1 x_{24}),  \\
&& \Phi_{134}\left(x_{13},x_{14},x_{34}\right)=(T_4 x_{13},T_3 x_{14},T_1 x_{34}), \\
&& \Phi_{234}\left(x_{23},x_{24},x_{34}\right)=(T_4 x_{23},T_3 x_{24},T_2x_{34}). 
\end{eqnarray*}
At the second stage, the map is applied to the other four 3D cubes of the 4D cube:
\begin{eqnarray*}
&& \Phi_{123}\left(T_4 x_{12},T_4x_{13},T_4x_{23}\right)=(T_4 (T_3  x_{12}),T_4 (T_2\, x_{13}),T_4(T_1  x_{23})),  \\
&& \Phi_{124}\left(T_3 x_{12},T_3x_{14},T_3x_{24}\right)=(T_3 (T_4  x_{12}),T_3 (T_2\, x_{14}),T_3(T_1  x_{24})), \\
&& \Phi_{134}\left(T_2 x_{13},T_2x_{14},T_2x_{34}\right)=(T_2 (T_4  x_{13}),T_2 (T_3\, x_{14}),T_2(T_1  x_{34})), \\
&& \Phi_{234}\left(T_1 x_{23},T_1x_{24},T_1x_{34}\right)=(T_1 (T_4  x_{23}),T_1 (T_3\, x_{24}),T_1(T_2  x_{34})). 
\end{eqnarray*}
Now, 4D consistency of $\Phi_{ijk}$ means that the following six equations are identically satisfied
for arbitrary initial data:
\beq \label{eq: 4D consistency}
T_\ell(T_k x_{ij})=T_k(T_\ell x_{ij}), \quad i<j,\quad  \{i,j,k,\ell\}=\{1,2,3,4\}.
\eeq

In this setting, the two most celebrated 4D consistent maps are the symmetric discrete Darboux system,
\begin{equation} \label{eq: Darboux shift}
T_k x_{ij}=\frac{x_{ij}+x_{ik}x_{jk}}{\sqrt{1-x_{ik}^2}\sqrt{1-x_{jk}^2}},
\end{equation}
(where one assumes that $x_{ij}=x_{ji}$), and the star-triangle map,
\begin{equation} \label{eq: star-triang shift}
T_k x_{ij}=-\frac{x_{ij}}{x_{ij}x_{jk}+x_{jk}x_{ki}+x_{ki}x_{ij}},
\end{equation}
(where one assumes that $x_{ij}=-x_{ji}$).

\section{Statement of the classification problem}
\label{sec3}

We consider maps (\ref{eq: map}) given by (formal or convergent) series of the following kind:
\beq \label{eq: map series}
T_k x_{ij}=x_{ij} + \sum_{m=2}^\infty A_{ij ; \, k}^{(m)}(x_{ij},x_{ik},x_{jk}),
\eeq
where $A_{ij;\, k}^{(m)}$ are homogeneous polynomials of degree $m$ of their respective arguments. Each time when the double indices are associated with a 2D coordinate plane (i.e., in notations like $x_{ij}$, $A_{ij;\, k}^{(m)}$) we assume that the two indices are permutable, that is, $x_{ij}=x_{ji}$, $A_{ij;\, k}^{(m)}=A_{ji;\, k}^{(m)}$, etc. Note that the Darboux system (\ref{eq: Darboux shift}) belongs to this class, while the star-triangle map (\ref{eq: star-triang shift}) does not  for two reasons: first, the skew-symmetry assumption $x_{ij}=-x_{ji}$ is enforced, and second, it is not close to identity.

We will classify 4D consistent maps of type (\ref{eq: map series}).
We perform the classification modulo the group of admissible transformations generated by the point (coordinate-wise) transformations
\beq \label{eq: gauge point}
x_{ij} \mapsto x_{ij}+ \sum_{m=2}^\infty b_{ij}^{(m)} x_{ij}^m,
\eeq
given by (formal or convergent) series, and by the scaling transformations
\beq \label{eq: gauge scale}
x_{ij} \mapsto  c_{ij} x_{ij}.
\eeq
Here $b_{ij}^{(m)}$ and $c_{ij}$ are arbitrary constants. Indeed, transformations from the group generated by (\ref{eq: gauge point}) and (\ref{eq: gauge scale}) leave the form of $\Phi_{ijk}$ invariant.

Our first aim is to determine quadratic polynomials $A_{ij;\, k}^{(2)}$ which can appear as leading terms in 4D consistent maps satisfying (\ref{eq: 4D consistency}). All lemmas in the present section are devoted to such a characterization.

\begin{lemma}
Polynomials $A_{ij;\, k}^{(2)}$ $(i,j,k\in\{1,2,3,4\})$  of 4D consistent maps (\ref{eq: map series}) satisfy the following system of six nonlinear partial differential equations (in which $\{i,j,k,\ell\}=\{1,2,3,4\}$):
\beq \label{eq. A2}
 \frac{\partial A_{ij;\, k}^{(2)}}{\partial x_{ij}} A_{ij;\, \ell}^{(2)} +
 \frac{\partial A_{ij;\, k}^{(2)}}{\partial x_{ik}} A_{ik;\, \ell}^{(2)} +
 \frac{\partial A_{ij;\, k}^{(2)}}{\partial x_{jk}} A_{jk;\, \ell}^{(2)} =
 \frac{\partial A_{ij;\, \ell}^{(2)}}{\partial x_{ij}} A_{ij;\, k}^{(2)} +
 \frac{\partial A_{ij;\, \ell}^{(2)}}{\partial x_{i\ell}} A_{i\ell;\, k}^{(2)} +
 \frac{\partial A_{ij;\, \ell}^{(2)}}{\partial x_{j\ell}} A_{j\ell;\, k}^{(2)}.
\eeq
Here and below, if the arguments of $A_{ij;\,k}^{(m)}$ are not written explicitly, they are supposed to be $(x_{ij},x_{ik},x_{jk})$.
\end{lemma}

\benpf We compare the terms of degree 3 in $T_\ell(T_k x_{ij})= T_k(T_\ell x_{ij})$. We have:
\begin{align*}
& T_k x_{ij}=x_{ij}+A_{ij; \, k}^{(2)}(x_{ij},x_{ik},x_{jk})+A_{ij; \, k}^{(3)}(x_{ij},x_{ik},x_{jk})+\ldots, \\
& T_\ell x_{ij}=x_{ij}+A_{ij;\,\ell}^{(2)}(x_{ij},x_{i\ell},x_{j\ell})+A_{ij;\,\ell}^{(3)}(x_{ij},x_{i\ell},x_{j\ell})+\ldots,
\end{align*}
so that, taking into account only terms relevant for the expansion up to degree 3,
\begin{align*}
 T_\ell(T_k x_{ij})= & \; x_{ij}+A_{ij;\,\ell}^{(2)}+A_{ij;\,\ell}^{(3)}+\ldots\\
 & +A_{ij;\,k}^{(2)}\left(x_{ij} + A_{ij;\,\ell}^{(2)}+\ldots,x_{ik}+A_{ik;\,\ell}^{(2)}+\ldots,x_{jk} + A_{jk;\,\ell}^{(2)}+\ldots \right)\\
 & +A_{ij;\,k}^{(3)}\left(x_{ij}+\ldots,x_{ik}+\ldots,x_{jk}+\ldots \right)+\ldots\\
 =& \; x_{ij}+A_{ij;\,\ell}^{(2)}+A_{ij;\,k}^{(2)}+A_{ij;\,\ell}^{(3)}+A_{ij;\,k}^{(3)}\\
  & +\frac{\partial A_{ij;\, k}^{(2)}}{\partial x_{ij}} A_{ij;\, \ell}^{(2)}
    +\frac{\partial A_{ij;\, k}^{(2)}}{\partial x_{ik}}A_{ik;\, \ell}^{(2)}
    +\frac{\partial A_{ij;\, k}^{(2)}}{\partial x_{jk}}A_{jk;\, \ell}^{(2)}+\ldots,
\end{align*}
and similarly,
\begin{align*}
 T_\ell(T_k x_{ij})= & \; x_{ij}+A_{ij;\,k}^{(2)}+A_{ij;\,k}^{(3)}+\ldots\\
 & +A_{ij;\,\ell}^{(2)}\left(x_{ij} + A_{ij;\,k}^{(2)}+\ldots,x_{i\ell}+A_{i\ell;\,k}^{(2)}+\ldots,x_{j\ell} + A_{j\ell;\,k}^{(2)}+\ldots \right)\\
 & +A_{ij;\,\ell}^{(3)}\left(x_{ij}+\ldots,x_{i\ell}+\ldots,x_{j\ell}+\ldots \right)+\ldots\\
 = & \; x_{ij}+A_{ij;\,k}^{(2)}+A_{ij;\,\ell}^{(2)}+A_{ij;\,k}^{(3)}+A_{ij;\,\ell}^{(3)}\\
 &  +\frac{\partial A_{ij;\, \ell}^{(2)}}{\partial x_{ij}} A_{ij;\, k}^{(2)}
    +\frac{\partial A_{ij;\, \ell}^{(2)}}{\partial x_{i\ell}}A_{i\ell;\, k}^{(2)}
    +\frac{\partial A_{ij;\, \ell}^{(2)}}{\partial x_{j\ell}}A_{j\ell;\, k}^{(2)}+\ldots.
\end{align*}
Now equations (\ref{eq. A2}) follow directly by comparing the terms of degree 3 in $T_\ell(T_k x_{ij})$ and $T_\ell(T_k x_{ij})$.
\endpf
\smallskip

Next, we analyze the system (\ref{eq. A2}) for polynomials $A_{ij;\, k}^{(2)}$. We use the following notation for their coefficients:
\beq
A_{ij;\, k}^{(2)}(x_{ij},x_{ik},x_{jk}) = \alpha_{ij; \, k} x_{ik} x_{jk} +
 \beta_{ij; \, k}^{(i)} x_{ij} x_{ik} +
  \beta_{ij; \, k}^{(j)} x_{ij} x_{jk} +
\lambda_{ij; \, k} x_{ij}^2 +
\mu_{ij; \, k}^{(i)} x_{ik}^2 +
\mu_{ij; \, k}^{(j)} x_{jk}^2 . \label{eq: defA2}
\eeq

\begin{lemma}
For polynomials $A_{ij;\, k}^{(2)}$ $(i,j,k\in\{1,2,3,4\})$ of the form  (\ref{eq: defA2}) satisfying system
(\ref{eq. A2}), we have
\beq
\alpha_{ij; \, \ell} \lambda_{ij; \, k}=0, \label{cond1}
\eeq
for all $\{i,j,k,\ell\}=\{1,2,3,4\}$.
\label{lemma2}
\end{lemma}

\benpf We substitute (\ref{eq: defA2}) into the set of differential equations (\ref{eq. A2}) thus obtaining a system of cubic polynomial equations which must be identically satisfied. The monomial $x_{ij}x_{i\ell}x_{j\ell}$ can only come from the first term on the left-hand side, and the vanishing of its coefficient results in condition (\ref{cond1}).
\endpf

We now deliberately choose to consider only the main two branches of conditions (\ref{cond1}), namely
\begin{enumerate}

\item[(I)] $\alpha_{ij; \, k}\neq 0$ and $\lambda_{ij; \, k}=0$ for all $i,j,k\in\{1,2,3,4\}$.

\item[(II)] $\alpha_{ij; \, k}= 0$ and $\lambda_{ij; \, k}\neq0$ for all $i,j,k\in\{1,2,3,4\}$.

\end{enumerate}
Accordingly we have the claims contained in Lemmas
\ref{lemma3} and \ref{lemma4}.

\begin{lemma} In case {\rm{(I)}},
polynomials $A_{ij;\, k}^{(2)}$ $(i,j,k\in\{1,2,3,4\})$ of the form  (\ref{eq: defA2}) satisfy system (\ref{eq. A2}) if and only if
\beq \label{case I}
\beta_{ij; \, k}^{(i)}=0, \qquad \mu_{ij; \, k}^{(i)}=0, \qquad
\alpha_{ij; \, k}= \frac{c_{ik} c_{jk}}{c_{ij}},
\eeq
for all $i,j,k\in\{1,2,3,4\}$, where $c_{ij}$ are arbitrary constants. Modulo gauge transformations, we can assume
$$
A_{ij;\, k}^{(2)} = x_{ik} x_{jk}
$$
for all $i,j,k\in\{1,2,3,4\}$

\label{lemma3}
\end{lemma}

\benpf Assume $\alpha_{ij; \, k}\neq 0$ and $\lambda_{ij; \, k}=0$ for all $i,j,k\in\{1,2,3,4\}$ and substitute (\ref{eq: defA2}) into the set of differential equations (\ref{eq. A2}). Collecting coefficients of monomials $x_{i\ell}x_{j\ell}x_{ik}$, we find
$
\alpha_{ij; \, \ell} (\beta_{ij; \, k}^{(i)}-\beta_{i\ell; \, k}^{(i)})=0,
$
which gives
\beq \label{beta1}
 \beta_{ij; \, k}^{(i)}=\beta_{i\ell; \, k}^{(i)}
\eeq
for all $\{i,j,k,\ell\}=\{1,2,3,4\}$. Similarly, collecting coefficients of monomials $x_{i\ell}x_{j\ell}x_{k\ell}$, we find
$
\alpha_{ij; \, \ell} (\beta_{i\ell; \, k}^{(\ell)}+ \beta_{j\ell; \, k}^{(\ell)})=0,
$
which gives
\beq \label{beta2}
\beta_{i\ell; \, k}^{(\ell)}=- \beta_{j\ell; \, k}^{(\ell)}
\eeq
for all $\{i,j,k,\ell\}=\{1,2,3,4\}$. Combining conditions (\ref{beta1}) and (\ref{beta2}) gives $\beta_{ij; \, k}^{(i)}=0$ for all $i,j,k\in\{1,2,3,4\}$. Collecting coefficients of monomials $x_{i\ell}^2x_{jk}$, we find
$
\mu_{ik; \, \ell}^{(i)} \alpha_{ij; \, k}=0,
$
which gives the remaining conditions $\mu_{ij; \, k}^{(i)}=0$ for all $i,j,k\in\{1,2,3,4\}$. We are now left with
\beq \label{A}
A_{ij;\, k}^{(2)} = \alpha_{ij;\, k}x_{ik} x_{jk}.
\eeq
Substituting this into (\ref{eq. A2}) we find the following conditions for the coefficients $\alpha_{ij;\, k}$:
\beq \label{cond alpha}
\alpha_{ik;\, \ell}\alpha_{ij;\, k}= \alpha_{j\ell;\, k}\alpha_{ij;\, \ell} \quad \Leftrightarrow \quad
\alpha_{jk;\, \ell}\alpha_{ij;\, k}= \alpha_{i\ell;\, k}\alpha_{ij;\, \ell}.
\eeq
To solve conditions (\ref{cond alpha}), we re-write them as
$$
\frac{\alpha_{ij;\, \ell}}{\alpha_{ij;\, k}}=\frac{\alpha_{ik;\, \ell}}{\alpha_{j\ell;\, k}}=\frac{\alpha_{jk;\, \ell}}{\alpha_{i\ell;\, k}}.
$$
The second equality here implies:
$$
\alpha_{ik;\, \ell}\alpha_{i\ell;\, k}=\alpha_{jk;\, \ell}\alpha_{j\ell;\, k}=: c_{k\ell}^2,
$$
since this quantity does not depend on $i$, $j$. As a consequence, we find:
$$
\alpha_{ij; \, k}^2= \frac{\alpha_{ij;\, k}\alpha_{jk;\, i}\cdot \alpha_{ij;\, k}\alpha_{ik;\, j}}{\alpha_{jk;\, i}\alpha_{ik;\, j}}=
\frac{c_{ik}^2 c_{jk}^2}{c_{ij}^2},
$$
which coincides with $\alpha_{ij;\, k}=c_{ik}c_{jk}/c_{ij}$ from (\ref{case I}). It is immediately verified that this necessary condition is also sufficient for (\ref{cond alpha}). The gauge transformation  (\ref{eq: gauge scale}) brings a map with the quadratic terms (\ref{A})  with $\alpha_{ij;\, k}=c_{ik}c_{jk}/c_{ij}$ to a similar map with $\alpha_{ij;\, k}=1$.
\endpf

\begin{lemma} In case {\rm{(II)}} polynomials $A_{ij;\, k}^{(2)}$ of the form  (\ref{eq: defA2}) satisfy system (\ref{eq. A2}) if and only if
$$
\beta_{ij; \, k}^{(i)}=0, \qquad \mu_{ij; \, k}^{(i)}=0,
$$
so that
$$
A_{ij;\, k}^{(2)} = \lambda_{ij;\, k}x_{ij}^2
$$
for all $i,j,k\in\{1,2,3,4\}$.
\label{lemma4}
\end{lemma}

\benpf Assume $\alpha_{ij; \, k}=0$ and $\lambda_{ij; \, k}\neq0$ for all $i,j,k\in\{1,2,3,4\}$ and substitute (\ref{eq: defA2}) into the set of differential equations (\ref{eq. A2}). Collecting coefficients of monomials $x_{ij}^2x_{ik}$, we find
$
\lambda_{ij; \, \ell} \beta_{ij; \, k}^{(i)}=0,
$
which gives $\beta_{ij; \, k}^{(i)}=0$ for all $i,j,k\in\{1,2,3,4\}$. Next, collecting coefficients of monomials $x_{i\ell}^2x_{ij}$, we find
$
\lambda_{ij; \, k} \mu_{ij; \, \ell}^{(i)}=0,
$
which gives the remaining conditions $\mu_{ij; \, k}^{(i)}=0$ for all $i,j,k\in\{1,2,3,4\}$. We are left with
$
A_{ij;\, k}^{(2)} = \lambda_{ij;\, k}x_{ij}^2,
$
which satisfy (\ref{eq. A2}) without any further conditions on the coefficients $\lambda_{ij;\, k}$.
\endpf

\section{Classification for case (I)}

The main result of our paper is contained in the following Theorem.

\begin{theorem} \label{Th 1}
Any 4D consistent system of map $\Phi_{ijk}$ $(i,j,k\in\{1,2,3,4\})$ of type (\ref{eq: map series}) with
$$
A_{ij; \, k}^{(2)}= x_{ik} x_{jk}
$$
is equivalent to the symmetric Darboux system (\ref{eq: Darboux shift}) modulo admissible transformations of type (\ref{eq: gauge point}).
\end{theorem}

\benpf
For any $m\ge 2$, the terms of degree $m+2$ in the 4D consistency condition (\ref{eq: 4D consistency}) yield the following recursive system of equations for the polynomials $A_{ij;\,k}^{(m+1)}$:
\begin{align}
& x_{ik} A_{jk;\,\ell}^{(m+1)}+
x_{jk} A_{ik;\, \ell}^{(m+1)}+
x_{i\ell} x_{j\ell}\frac{\partial A_{ij;\, k}^{(m+1)}}{\partial x_{ij}}+
x_{i\ell} x_{k \ell}\frac{\partial A_{ij;\, k}^{(m+1)}}{\partial x_{ik}}+
x_{j\ell} x_{k\ell}\frac{\partial A_{ij;\, k}^{(m+1)}}{\partial x_{jk}} \label{eq: 1} \\
&-x_{i\ell} A_{j\ell;\,k}^{(m+1)}-
x_{j\ell} A_{i\ell;\, k}^{(m+1)}-
x_{ik} x_{jk}\frac{\partial A_{ij;\, \ell}^{(m+1)}}{\partial x_{ij}}-
x_{ik} x_{k \ell}\frac{\partial A_{ij;\, \ell}^{(m+1)}}{\partial x_{i\ell}}-
x_{jk} x_{k\ell}\frac{\partial A_{ij;\, \ell}^{(m+1)}}{\partial x_{j\ell}}\!=\!F\left(\!A_{\alpha\beta;\, \gamma}^{(s)}\!\right), \nn
\end{align}
where $F$ is a polynomial function of the $A_{\alpha\beta;\, \gamma}^{(s)}$'s with $s\leq m$. This is a system of six partial differential equations for twelve unknown polynomials $A_{ij;\, k}^{(m+1)}$, which can be re-formulated as a system of linear algebraic equations for the coefficients of these polynomials. We know that (\ref{eq: 1}) admits at least one particular solution (corresponding to the symmetric Darboux map), and we are going to study the general solution. It is obtained by adding an arbitrary solution of the corresponding homogeneous system:
\begin{align}
& x_{ik} A_{jk;\,\ell}^{(m+1)}+ x_{jk} A_{ik;\, \ell}^{(m+1)}+
x_{i\ell} x_{j\ell}\frac{\partial A_{ij;\, k}^{(m+1)}}{\partial x_{ij}}+
x_{i\ell} x_{k \ell}\frac{\partial A_{ij;\, k}^{(m+1)}}{\partial x_{ik}}+
x_{j\ell} x_{k\ell}\frac{\partial A_{ij;\, k}^{(m+1)}}{\partial x_{jk}} \label{eq: 12} \\
& -x_{i\ell} A_{j\ell;\,k}^{(m+1)}-
x_{j\ell} A_{i\ell;\, k}^{(m+1)}-
x_{ik} x_{jk}\frac{\partial A_{ij;\, \ell}^{(m+1)}}{\partial x_{ij}}-
x_{ik} x_{k \ell}\frac{\partial A_{ij;\, \ell}^{(m+1)}}{\partial x_{i\ell}}-
x_{jk} x_{k\ell}\frac{\partial A_{ij;\, \ell}^{(m+1)}}{\partial x_{j\ell}}=0. \nn
\end{align}

\begin{lemma} \label{Lemma 1}
The general solution of the homogeneous system (\ref{eq: 12}) is given by
\beq
A_{ij; \, k}^{(m+1)} =
 x_{ik} x_{jk} \left( m b_{ij} x_{ij}^{m-1} - b_{ik}x_{ik}^{m-1} - b_{jk}x_{jk}^{m-1}\right),
 \label{Alemma}
\eeq
where $b_{ij}$ are arbitrary constants.
\label{lemma5}
\end{lemma}

It turns out that this freedom is exactly compensated  by means of admissible gauge transformations:

\begin{lemma}
Given maps $\Phi_{ijk}$ $(i,j,k\in\{1,2,3,4\})$
of type (\ref{eq: map series}), a change of variables
$$
x_{ij}\mapsto x_{ij} + b_{ij} x_{ij}^{m}
$$
with $m\ge 2$ does not change the polynomials $A_{ij;\,k}^{(s)}$, with $s\leq m$,
while the polynomials $A_{ij;\,k}^{(m+1)}$ get shifted by
$$
x_{ik} x_{jk}\left( - m b_{ij} x_{ij}^{m-1}+
 b_{ik} x_{ik}^{m-1} + b_{jk} x_{jk}^{m-1}\right).
$$
\label{lemma6}
\end{lemma}

Theorem \ref{Th 1} is an immediate consequence of Lemmas \ref{lemma5} and \ref{lemma6}.
\endpf

\smallskip

It remains to prove Lemmas \ref{lemma5} and \ref{lemma6}. The second one is proved by a straightforward computation.

\bigskip

\noindent {\textbf{\emph{Proof}}} ({\emph{of Lemma \ref{lemma5}}}).
We consider in (\ref{eq: 12}) terms of a  low multidegree with respect to $x_{i\ell}$, $x_{j\ell}$, $x_{k\ell}$, belonging to three distinct classes:

\begin{enumerate}

\item[(a)] Terms of multidegree $(1,1,0)$ with respect to $x_{i\ell}$, $x_{j\ell}$, $x_{k\ell}$, and of degree $m-1$ with respect to $x_{ij}$, $x_{ik}$, $x_{jk}$.

\item[(b)] Terms of multidegree $(1,0,1)$ with respect to $x_{i\ell}$, $x_{j\ell}$, $x_{k\ell}$, and of degree $m-1$ with respect to $x_{ij}$, $x_{ik}$, $x_{jk}$.

\item[(c)] Terms of multidegree $(0,1,1)$ with respect to $x_{i\ell}$, $x_{j\ell}$, $x_{k\ell}$, and of degree $m-1$ with respect to $x_{ij}$, $x_{ik}$, $x_{jk}$.

\end{enumerate}

\noindent Terms belonging to the class (a)
can come from the following part of (\ref{eq: 12}):
\[
x_{i\ell} x_{j\ell}\frac{\partial A_{ij;\, k}^{(m+1)}}{\partial x_{ij}}-
x_{i\ell} A_{j\ell;\,k}^{(m+1)}-
x_{j\ell} A_{i\ell;\, k}^{(m+1)}-
x_{ik} x_{jk}\frac{\partial A_{ij;\, \ell}^{(m+1)}}{\partial x_{ij}},
\]
which, upon cancelation of the common factor $x_{i\ell} x_{j\ell}$, leads to
\beq \label{eq: proof1 1}
\frac{\partial A_{ij;\, k}^{(m+1)}}{\partial x_{ij}}= p_1 x_{jk}^m +p_2 x_{ik}^m +p_3 x_{ik} x_{jk} x_{ij}^{m-2},
\eeq
where
\begin{itemize}
\item $p_1$ is the coefficient by $x_{j\ell}x_{jk}^m$ in $A_{j\ell;\,k}^{(m+1)}$,
\item $p_2$ is the coefficient by $x_{i\ell}x_{ik}^m$ in $A_{i\ell;\,k}^{(m+1)}$,
\item $p_3$ is the coefficient by $x_{i\ell}x_{j\ell}x_{ij}^{m-2}$ in $\partial A_{ij;\,\ell}^{(m+1)}/\partial x_{ij}$.
\end{itemize}

\noindent Terms belonging to the class (b)
can come from the following part of (\ref{eq: 12}):
\[
x_{jk} A_{ik;\, \ell}^{(m+1)}+
x_{i\ell} x_{k \ell}\frac{\partial A_{ij;\, k}^{(m+1)}}{\partial x_{ik}}-x_{i\ell} A_{j\ell;\,k}^{(m+1)}-
x_{ik} x_{k \ell}\frac{\partial A_{ij;\, \ell}^{(m+1)}}{\partial x_{i\ell}}-
x_{jk} x_{k\ell}\frac{\partial A_{ij;\, \ell}^{(m+1)}}{\partial x_{j\ell}},
\]
which, upon cancelation of the common factor $x_{i\ell} x_{k\ell}$, leads to
\beq\label{eq: proof1 2}
\frac{\partial A_{ij;\, k}^{(m+1)}}{\partial x_{ik}}=
- p_4 x_{jk} x_{ik}^{m-1}+ p_5 x_{jk}^m +
p_6 x_{ik} x_{ij}^{m-1}+
p_7 x_{jk} x_{ij}^{m-1},
\eeq
where
\begin{itemize}
\item $p_4$ is the coefficient by $x_{i\ell}x_{k\ell}x_{ik}^{m-1}$ in $A_{ik;\, \ell}^{(m+1)}$,
\item $p_5$ is the coefficient by $x_{k\ell}x_{jk}^m$ in $A_{j\ell;\,k}^{(m+1)}$,
\item $p_6$ is the coefficient by $x_{i\ell}x_{ij}^{m-1}$ in $\partial A_{ij;\, \ell}^{(m+1)}/\partial x_{i\ell}$,
\item $p_7$ is the coefficient by $x_{i\ell}x_{ij}^{m-1}$ in $\partial A_{ij;\, \ell}^{(m+1)}/\partial x_{j\ell}$.
\end{itemize}

\noindent Finally, terms belonging to the class (c)
can come from the following part of (\ref{eq: 12}):
\[
x_{ik} A_{jk;\,\ell}^{(m+1)}+
x_{j\ell} x_{k\ell}\frac{\partial A_{ij;\, k}^{(m+1)}}{\partial x_{jk}}-
x_{j\ell} A_{i\ell;\, k}^{(m+1)}-
x_{ik} x_{k \ell}\frac{\partial A_{ij;\, \ell}^{(m+1)}}{\partial x_{i\ell}}-
x_{jk} x_{k\ell}\frac{\partial A_{ij;\, \ell}^{(m+1)}}{\partial x_{j\ell}},
\]
which, upon cancelation of the common factor $x_{j\ell} x_{k\ell}$, leads to
\beq\label{eq: proof1 3}
\frac{\partial A_{ij;\, k}^{(m+1)}}{\partial x_{jk}}=
- p_8 x_{ik} x_{jk}^{m-1}
+ p_9 x_{ik}^m
+ p_{10} x_{ik} x_{ij}^{m-1}
+ p_{11} x_{jk} x_{ij}^{m-1},
\eeq
where
\begin{itemize}
\item $p_8$ is the coefficient by $x_{j\ell}x_{k\ell}x_{jk}^{m-1}$ in $A_{jk;\,\ell}^{(m+1)}$,
\item $p_9$ is the coefficient by $x_{k\ell}x_{ik}^m$ in $A_{i\ell;\, k}^{(m+1)}$,
\item $p_{10}$ is the coefficient by $x_{j\ell}x_{ij}^{m-1}$ in $\partial A_{ij;\, \ell}^{(m+1)}/\partial x_{i\ell}$,
\item $p_{11}$ is the coefficient by $x_{j\ell}x_{ij}^{m-1}$ in $\partial A_{ij;\, \ell}^{(m+1)}/\partial x_{j\ell}$.
\end{itemize}
\medskip

Cross-differentiation of expressions in (\ref{eq: proof1 1}), (\ref{eq: proof1 2}), (\ref{eq: proof1 3}) leads to:
\begin{eqnarray*}
\frac{\partial^2 A_{ij;\,k}^{(m+1)}}{\partial x_{ij}\partial x_{ik}}
  & = & mp_2 x_{ik}^{m-1}+p_3 x_{jk}x_{ij}^{m-2} \\
  & = & (m-1)p_6x_{ik}x_{ij}^{m-2}+(m-1)p_7x_{jk}x_{ij}^{m-2},\\
\frac{\partial^2 A_{ij;\,k}^{(m+1)}}{\partial x_{ij}\partial x_{jk}}
  & = & mp_1x_{jk}^{m-1}+p_3x_{ik}x_{ij}^{m-2} \\
  & = & (m-1)p_{10}x_{ik}x_{ij}^{m-2}+(m-1)p_{11}x_{jk}x_{ij}^{m-2},\\
\frac{\partial^2 A_{ij;\,k}^{(m+1)}}{\partial x_{ik}\partial x_{jk}}
  & = & -p_4 x_{ik}^{m-1}+mp_5 x_{jk}^{m-1}+p_7x_{ij}^{m-1} \\
  & = & -p_8x_{jk}^{m-1}+mp_9x_{ik}^{m-1}+p_{10}x_{ij}^{m-1}.\
\end{eqnarray*}
Thus, we find the conditions
$$
p_1=p_2=p_6=p_{11}=0,
$$
$$
p_3=(m-1)p_7=(m-1)p_{10},
$$
$$
p_4=-mp_9,\quad p_8=-mp_5.
$$
Taking into account the above conditions
we can write our preliminary results as follows:
\begin{eqnarray}
\frac{\partial A_{ij;\, k}^{(m+1)}}{\partial x_{ij}} & = &
(m-1)p_{10} x_{ik} x_{jk} x_{ij}^{m-2},
\label{eq: proof1 4}\\
\frac{\partial A_{ij;\, k}^{(m+1)}}{\partial x_{ik}} & = &
mp_{9} x_{jk} x_{ik}^{m-1}+ p_5 x_{jk}^m +p_{10} x_{jk} x_{ij}^{m-1},
\label{eq: proof1 5}\\
\frac{\partial A_{ij;\, k}^{(m+1)}}{\partial x_{jk}} & = &
 mp_5 x_{ik} x_{jk}^{m-1} + p_9 x_{ik}^m + p_{10} x_{ik} x_{ij}^{m-1},
 \label{eq: proof1 6}
\end{eqnarray}
where
\begin{itemize}
\item $(m-1)p_{10}$ is the coefficient by $x_{i\ell}x_{j\ell}x_{ij}^{m-2}$ in $\partial A_{ij;\,\ell}^{(m+1)}/\partial x_{ij}$,
\item $p_{10}$ is the coefficient by $x_{i\ell}x_{ij}^{m-1}$ in $\partial A_{ij;\, \ell}^{(m+1)}/\partial x_{j\ell}$, and at the same time the coefficient by $x_{j\ell}x_{ij}^{m-1}$ in $\partial A_{ij;\, \ell}^{(m+1)}/\partial x_{i\ell}$,
\item $p_5$ is the coefficient by $x_{k\ell}x_{jk}^m$ in $A_{j\ell;\,k}^{(m+1)}$,
\item $-mp_5$ is the coefficient by $x_{j\ell}x_{k\ell}x_{jk}^{m-1}$ in $A_{jk;\,\ell}^{(m+1)}$,
\item $p_9$ is the coefficient by $x_{k\ell}x_{ik}^m$ in $A_{i\ell;\, k}^{(m+1)}$,
\item $-mp_9$ is the coefficient by $x_{i\ell}x_{k\ell}x_{ik}^{m-1}$ in $A_{ik;\, \ell}^{(m+1)}$.
\end{itemize}
Now from equations (\ref{eq: proof1 4})--(\ref{eq: proof1 6}) we find:
\begin{equation}\label{eq: proof1 res}
A_{ij;\,k}^{(m+1)}=p_5 x_{ik}x_{jk}^m+p_9 x_{jk}x_{ik}^m+p_{10}x_{ik}x_{jk}x_{ij}^{m-1},
\end{equation}
where
\begin{itemize}
\item $p_{10}$ is the coefficient by $x_{i\ell}x_{j\ell}x_{ij}^{m-1}$ in $A_{ij;\, \ell}^{(m+1)}$,
\item $-mp_5$ is the coefficient by $x_{j\ell}x_{k\ell}x_{jk}^{m-1}$ in $A_{jk;\,\ell}^{(m+1)}$,
\item $-mp_9$ is the coefficient by $x_{i\ell}x_{k\ell}x_{ik}^{m-1}$ in $A_{ik;\, \ell}^{(m+1)}$.
\end{itemize}
From the characterization of $p_{10}$ it follows that this coefficient depends on $i,j$ only, being independent of $k,\ell$. Thus, we can set
$
p_{10}=mb_{ij}.
$
Then the characterizations of $-mp_5$ and $-mp_9$ yield that $p_5=-b_{jk}$, $p_9=-b_{ik}$.
This allows us to finally re-write (\ref{eq: proof1 res}) in the form (\ref{Alemma}), which proves Lemma \ref{lemma5}.
\endpf

\section{Classification for case (II)}

In the case (II) we only have trivial 4D consistent maps.

\begin{theorem} \label{Th 2}
Any 4D consistent system of map $\Phi_{ijk}$ $(i,j,k\in\{1,2,3,4\})$ of type (\ref{eq: map series}) with
\beq \label{Th 2 A}
A_{ij; \, k}^{(2)}= \lambda_{ij;\, k}x_{ij}^2,
\eeq
$\lambda_{ij;\, k}\neq 0$, is equivalent to
\beq \label{Th 2 f}
T_k x_{ij} = f_{ij; \, k} (x_{ij}),
\eeq
where $f_{ij;\, k}$ are univariate functions such that
\beq \label{comm}
f_{ij; \, k} \circ f_{ij; \, \ell}=f_{ij; \, \ell} \circ f_{ij; \, k}
\eeq
for all $i,j,k,\ell\in\{1,2,3,4\}$.
\end{theorem}

\benpf
We prove by induction that in the expression (\ref{eq: map series}) for $T_k x_{ij}$ each polynomial $A_{ij;\, k}^{(m)}$ of degree $m$  is in fact a monomial $\lambda_{ij;\, k}^{(m)}x_{ij}^m$ depending on $x_{ij}$ only. By assumption, this is true for $m=2$. We fix $m\ge 3$, assume the said property is fulfilled for all degrees less than $m$ and derive the same property for degree $m$. One computes that the terms of degree $m+1$ in $T_\ell(T_k x_{ij})-T_k(T_\ell x_{ij})$ are given by
\begin{align}\label{eq: aux}
& 2\lambda_{ij;\, k}x_{ij}A_{ij;\, \ell}^{(m)}+
\frac{\partial A_{ij;\, k}^{(m)}}{\partial x_{ij}}\lambda_{ij;\, \ell}x_{ij}^2 +
\frac{\partial A_{ij;\, k}^{(m)}}{\partial x_{ik}}\lambda_{ik;\, \ell}x_{ik}^2 +
\frac{\partial A_{ij;\, k}^{(m)}}{\partial x_{jk}}\lambda_{jk;\, \ell}x_{jk}^2 \nn \\
& -2\lambda_{ij;\, \ell}x_{ij}A_{ij;\, k}^{(m)}-
\frac{\partial A_{ij;\, \ell}^{(m)}}{\partial x_{ij}}\lambda_{ij;\, k}x_{ij}^2 -
\frac{\partial A_{ij;\, \ell}^{(m)}}{\partial x_{i\ell}}\lambda_{i\ell;\, k}x_{i\ell}^2 -
\frac{\partial A_{ij;\, \ell}^{(m)}}{\partial x_{j\ell}}\lambda_{j\ell;\, k}x_{j\ell}^2+px_{ij}^{m+1},
\end{align}
where the term $px_{ij}^{m+1}$ comes from $A_{ij;\, k}^{(s)}$, $A_{ij;\, \ell}^{(s)}$ with $s<m$. Now it is easy to show that $A_{ij;\, k}^{(m)}$ does not depend on $x_{ik}$, $x_{jk}$. Indeed, if the degree of $A_{ij;\, k}^{(m)}$ with respect to $x_{ik}$ would be $d\in[1,m]$ (with a non-vanishing leading coefficient), then the degree of the expression (\ref{eq: aux}) with respect to $x_{ik}$ would be $d+1$ (also with a non-vanishing leading coefficient), so this expression could not vanish identically.

We conclude that any 4D consistent system of maps $\Phi_{ijk}$ $(i,j,k\in\{1,2,3,4\})$ of type (\ref{eq: map series}) with (\ref{Th 2 A}) is equivalent to (\ref{Th 2 f}) for some functions $f_{ij;\, k}$. The 4D consistency condition is then nothing but that the requirement that the functions $f_{ij; \, k}$ commute as in (\ref{comm}).
\endpf

\section{Conclusions}

The present paper has been devoted to the classification of 3D maps of type (\ref{eq: map series}) which are 4D consistent. The results of our classification problem are contained in Theorems \ref{Th 1} and \ref{Th 2}. The most interesting finding is that the symmetric discrete Darboux system (\ref{eq: Darboux shift})
is the only 4D consistent map belonging to our Ansatz.

Future research will be devoted to the classification of 4D consistent 3D maps which are not a perturbation of the identity (thus including the star-triangle map (\ref{eq: star-triang shift})), as well as to extension into higher dimensions. Let us mention that no 4D map  which would be 5D consistent is presently known.

\section*{Acknowledgment}
This research is supported by the DFG Collaborative Research Center TRR 109 ``Discretization in Geometry and Dynamics''.

\end{document}